# Bayesian Modeling of Nonlinear Poisson Regression with Artificial Neural Networks

Hansapani Rodrigo[1] and Chris Tsokos[2]

[1] University of Texas Rio Grande Valley, Edinburg TX 78539, USA

**hansapani.rodrigo@utrgv.edu**

[2] University of South Florida, Tampa FL 33520, USA

**ctsokos@usf.edu**

**Abstract**. Being in the era of big data, modeling and prediction of count data have become significantly important in many fields including health, finance, social, etc. Although linear Poisson regression has been widely used to model count and rate data, it might not be always suitable as it cannot capture some inherent variability within complex data. In this study, we introduce a probabilistically driven nonlinear Poisson regression model with Bayesian artificial neural networks (ANN) to model count or rate data. This new nonlinear Poisson regression model developed with Bayesian ANN provides higher prediction accuracies over traditional Poisson or negative binomial regression models as revealed in our simulation and real data studies.

## 1. **Introduction**

Poisson regression is a form of regression analysis which is used to model count data [1]. This plays an important role in interdisciplinary research including health, finance, social, etc. For example, Poisson regression can be used to model the number of occurrences of mineral deposits [2], model number of insurance claims occurring in a given period [3], or to model the highway fatalities [4].

When developing a Poisson regression model, we assume that its mean is related to a function of covariates. More specifically, it assumes that the log-transformed outcomes are linearly related to the covariates. Nevertheless, this linearity assumption is not always appropriate for modelling count data. Another strong assumption related with Poisson distribution is that, it has identical mean and variance, which is most of the time count data do not adherence into. Any violation of this assumption in the model (known as the overdispersion) leads to significantly underestimated standard errors, and eventually provides misleading information on the significance of each covariate.

Over the last few decades, nonlinear modeling with artificial neural networks has gained an immense attraction due to their flexibility and high predictive performances. A significant number of researchers has contributed in developing nonlinear versions of generalized linear models with neural networks. Development of a nonlinear logistic regression model with an ANN can be considered as one of the pioneering study [5] in the statistical field. Following to that, C. M. B. Bishop in 2006 [6], has introduced a nonlinear multinomial logistic regression model using ANNs. Both of these models have been extensively applied for solving various interdisciplinary research problems [7, 8]. A nonlinear extension of ordinal logistic regression using ANN has been introduced in financial engineering by Mathieson *et al.* [9].

A nonlinear Poisson regression model with ANN has first been introduced by Fallah *et al.* [10] in 2009 utilizing the maximum likelihood (ML) approach. With this method, we can find an optimal set of weights by minimizing the error between the actual and the predicted outcomes. However, training ANN with ML tends to provide poor predictions due to its inherent problem of network overfitting which might lead to bias parameter estimations. The above model has been successfully used for predicting the cause-specific hazard of the breast cancer patients [11]. As per our knowledge, these are the only two studies which have contributed in developing a nonlinear Poisson regression model using ANN. In this study, we discuss a novel method of developing a nonlinear Poisson regression model with the Bayesian ANN. In fact, we introduce a Bayesian ANN for Poisson regression using a new hybrid Bayesian learning method which is based on the evidence procedure [12] and the hybrid Monte Carlo (HMC) [13] sampling. Although, Bayesian learning approaches with ANN have been used with regular regression [14], none of the existing studies have utilized them for count modeling. Use of Bayesian ANN in count modeling resulted in higher prediction accuracies as evident in our simulation and real data studies. Moreover, our approach allows researchers to use the automatic relevance determination prior (ARD) assess the importance of each covariate instead of p-values which might have affected due to the overdispersion. A successful real-world application of this new nonlinear Poisson regression model with Bayesian ANN can be found in the author's work [15].

The rest of the paper is organized as follows. In section 2, we introduce the Bayesian learning methods for ANNs with respect to count modeling including the details of the new hybrid Bayesian learning method for ANNs. In section 3, we present the results of our simulation and real data studies along with the convergence diagnostics checks for the hybrid Bayesian ANN model. The paper concludes with a discussion detailing the challenges and future use of the new ANN model.

## 2. Methods

### 2.1 Bayesian Artificial Neural Networks and Nonlinear Poisson Regression

An artificial neural network is an information processing archetype that is inspired by the biological neural networks systems, such as the human brain. They have been successfully applied in almost every field including engineering, computer science, medicine, etc. [7, 16–18]. The popularity of these models has increased mainly because of their flexibility associated with ANN modelling.

An ANN serves as a powerful tool for modeling nonlinear functions and non-additive effects. It has the strength of making predictions based on both individual attributable variables (covariates) and possible complex interactions among them. An ANN is organized as several interconnected layers; input, hidden and output, where each layer is a collection of artificial neurons and connections among these layers are made using weights (Figure 1). ANN follow a supervised learning approach where both inputs and outputs need to be fed into the network during the training phase. As a part of learning, the associated weights get adjusted in a way that the error between the actual and the predicted outcomes are minimized. The $k^{th}$ outcome of an ANN with has $d$ inputs, $M$ hidden and $K$ outputs nodes are given by the equation (1).

$$y_k(\boldsymbol{x}^n, \boldsymbol{w}) = g\left(\sum_{j=1}^{M} w_{kj}^{(2)} h\left(\sum_{i=1}^{d} w_{j1}^{(1)} x_i^n + b_j^{(1)}\right) + b_k^{(2)}\right) \tag{1}$$

Here $h$, and $g$ are the hidden and the output layer activation functions, respectively.

The training of an ANN can be done using either the maximum likelihood or the Bayesian methods. Bayesian neural networks provide a more intuitive approach for network training. A significant amount of research in this area has been conducted by David Mackay in 1992 [12, 19–21]. In the ML method, we find a single set of weight parameters by minimizing the error function. In contrast to that, in the Bayesian approach, a probability distribution is used to capture the uncertainties associated with the weight parameters [22]. Use of Bayesian learning in ANN provides several advantages over the ML method. It allows to use a relatively large number of regularization parameters while optimizing them during the training process. These regularization parameters have a natural interpretation in the Bayesian setting. Moreover, the ARD prior [13, 21, 23] helps to identify the relative importance of each covariate. The improved prediction accuracies can be obtained by creating network committees, i.e. by combining several ANN models. Error bars can be used to visualize the variations associated with the predictions.

## 2.2 Prior and Posterior Distributions

In this section, we discuss the Bayesian learning process of ANN in the context of nonlinear Poisson regression. The first step in Bayesian learning of ANN involves introducing a prior distribution for the weights. In this regard, we used a zero mean Gaussian prior of the form in equation (2),

$$p(\boldsymbol{w}|\boldsymbol{x}) = \frac{1}{Z_{\boldsymbol{w}}(\alpha)} exp^{\left(-\frac{\alpha}{2}w^T w\right)} = \frac{1}{Z_{\boldsymbol{w}}(\alpha)} exp^{\left(-\alpha E_p(\boldsymbol{w})\right)} \tag{2}$$

where $Z_{\boldsymbol{w}} = \left(\frac{2\pi}{\boldsymbol{\alpha}}\right)^{\frac{w}{2}}$. Here, $\boldsymbol{w}$ is the vector weights, $\boldsymbol{x}$ is the vector of inputs and $\boldsymbol{\alpha}$ is the hyperparameters of the prior distribution. As a part of Bayesian learning, we can optimize this hyperparameters (refer section 2.3). The Error term $E_{\boldsymbol{w}}$ is chosen to be ½ $\boldsymbol{w}^T\boldsymbol{w}$, as it penalizes the weights of large magnitudes and hence leads to a better generalization.

For a set of independent and identical count data $D = (t^1, ..., t^n)$, which follow a Poisson distribution with rate $\lambda^n$, the likelihood distribution can be derived as follows.

$$p(D|\boldsymbol{w}) = \prod_{n=1}^{N} p(t^n|\boldsymbol{x}^n, \boldsymbol{w}) = \prod_{n=1}^{N} \frac{e^{-\lambda^n}(\lambda^n)t^n}{t^n!}, t^n = 0,1,2, .... \tag{3}$$

where $p(t^n|\boldsymbol{x}^n, \boldsymbol{w})$ is the likelihood of data $t^n$. Our goal is to model the expected value of the Poison regression model, $\lambda^n$, using ANN with a hyperbolic tangent and an exponential activation function in their hidden and output layers, respectively. In order to do this, we first find the weight posterior distribution, $p(\boldsymbol{w}|D, \boldsymbol{x})$ using (2) and (3),

$$p(\boldsymbol{w}|D, \boldsymbol{x}) \propto p(D|\boldsymbol{w})p(\boldsymbol{w}|\boldsymbol{x}) \tag{4}$$

The regularized canonical error function of the Poisson regression model can be derived by taking the negative log-likelihood of the above posterior distribution,

$$E = S(\boldsymbol{w}) = \sum_{n=1}^{N} \left[ \{-y(\boldsymbol{x}^n, \boldsymbol{w}) + t^n \log(y(\boldsymbol{x}^n, \boldsymbol{w}))\} + \frac{\boldsymbol{\alpha}}{2} \boldsymbol{w}^T \boldsymbol{w} \right] = E_D + \boldsymbol{\alpha} E_w \tag{5}$$

As a part of the learning process, we minimize this error function and it can be done in two ways: using either the ML or the Bayesian approaches. In this study, we introduce a new hybrid Bayesian learning method based on the two existing Bayesian methods known as the evidence and the hybrid Monte Carlo as discussed in section 2.3. The posterior distribution of the weights, $p(\boldsymbol{w}|D, \boldsymbol{x})$ is used when making the predictions to a new set of covariates $\boldsymbol{x}^*$ when using the predictive distribution $p(t^*|\boldsymbol{x}^*, D)$

$$p(t^*|\boldsymbol{x}^*, D) = \int p(t^*|\boldsymbol{x}^*, \boldsymbol{w}) p(\boldsymbol{w}|D, \boldsymbol{x}) d\boldsymbol{w} \tag{6}$$

## 2.3 The Evidence Procedure and the Hybrid Monte Carlo Method

The evidence procedure is an iterative algorithm for determining the optimal weights and hyperparameters. Here, we present the details of the procedure very briefly. The weight posterior distribution in equation (4) can be rewritten by highlighting dependency of that on its hyperparameters,

$$p(\boldsymbol{w}|D, \boldsymbol{x}) = \int p(\boldsymbol{w}, \boldsymbol{\alpha}|D, \boldsymbol{x}) d\boldsymbol{\alpha} = \int p(\boldsymbol{w}, \boldsymbol{\alpha}, D, \boldsymbol{x}) p(\boldsymbol{\alpha}|D, \boldsymbol{x}) d\boldsymbol{\alpha} \tag{7}$$

Under the evidence procedure, we assume that the posterior density of the hyperparameters $p(\boldsymbol{\alpha}|D, \boldsymbol{x})$ is sharply peaked around the most probable values of the hyperparameter $\boldsymbol{\alpha}_{MAP}$. Therefore, using the Laplace approximation, we obtain,

$$p(\boldsymbol{w}|D, \boldsymbol{x}) \approx p(\boldsymbol{w}|\boldsymbol{\alpha}_{MAP}, D, \boldsymbol{x}) \int p(\boldsymbol{\alpha}|D, \boldsymbol{x}) d\boldsymbol{\alpha} \approx p(\boldsymbol{w}|\boldsymbol{\alpha}_{MAP}, D, \boldsymbol{x}) \tag{8}$$

Hence, prior to any other calculations, we need to find the value of $\boldsymbol{\alpha}_{MAP}$. The first step in the evidence procedure is to evaluate the posterior distribution of hyperparameter by approximating it with the most probable values of the hyperparameter. Once we found the $\boldsymbol{\alpha}_{MAP}$, we can approximate the regularized canonical error function using the second-order Taylor series expansion around the most probable weight vector $\boldsymbol{w}_{MAP}$ ,

$$S(\boldsymbol{w}) \approx S(\boldsymbol{w}_{MAP}) + \frac{1}{2} (\boldsymbol{w} - \boldsymbol{w}_{MAP})^T \boldsymbol{A} (\boldsymbol{w} - \boldsymbol{w}_{MAP}) \tag{9}$$

where $\boldsymbol{A} = \nabla\nabla S(\boldsymbol{w}_{MAP})$.

When $S(\boldsymbol{w})$ is at a given local minimum, we can re-estimate the hyperparameter $\alpha$ by,

$$\boldsymbol{\alpha}^{new} = \frac{\gamma}{2E_w} \tag{10}$$

where $\gamma = \sum_{i=1}^{W} \frac{\lambda_i}{\lambda_i + \alpha}$ and $\lambda_1, \ldots, \lambda_W$ are the eigen values of $\nabla\nabla E_D$. In the Evidence procedure, this process is repeated until we obtain the convergence. Finally, new predictions are made using equation 6. The evidence process searches for optimal parameters instead of integrating over all unknown parameters. Hence it is less computationally costly compared to other Bayesian approaches. This method has been applied in many applications effectively [24].

Unlike the evidence procedure which uses several approximations to get the weight posterior distribution and to optimize the hyperparameters, in HMC sampling method, the predictive distribution in equation 6 is approximated by a finite sum,

$$\langle p(t^*)\rangle = p(t^*|\boldsymbol{x}^*, D) \cong \frac{1}{N}\sum_{n=1}^{N} p(t^*|\boldsymbol{x}^*, \boldsymbol{w}_n), \tag{11}$$

where $\{\boldsymbol{w}_n\}$ represents a sample of weight vectors generated from the posterior distribution $p(\boldsymbol{w}|D,\boldsymbol{x})$. We also can obtain a statistical error estimate for our predictions by considering the variance of this statistic,

$$SE = \sqrt{\frac{\langle p(t^*)^2\rangle - (\langle p(t^*)\rangle)^2}{N}} \tag{12}$$

HMC method of sampling uses the information of gradients which makes it ideal for ANN modeling. Ideally, the accuracy of the above estimator does not depend on the dimensionality weight vector and hence higher prediction accuracies are expected with a relatively small number of samples. However, in reality, a large number of samples might require due to samples being not independent
A significant effort is needed in the process of selecting an appropriate informative prior along with the corresponding hyperparameter values.

## 2.4 New Hybrid Bayesian Learning Method for ANNs

In the HMC method, we need to generate several samples out of the posterior distribution of the weights in order to approximate the integral in equation (6) to make the predictions. The generation of these samples highly depends on the initial hyperparameter value of the weight posterior.

Prior to the HMC sampling, we can use the evidence procedure to optimize the hyperparameter value in the ANN model. This optimized hyperparameter value along with the weight parameters can then be used to generate the samples from the posterior distribution. This new approach, called the hybrid Bayesian, provides relatively high prediction accuracies, compared to HMC method alone.

Additionally, we can identify the relative importance of the covariates in the final ANN model. This can be achieved by integrating a separate hyperparameter to each covariate, representing the inverse variance of the prior distribution of the weights fanning out from that covariate [25]. The weights connected to the most relevant covariates are automatically set to small values and this is known as the ARD prior. Moreover, we can capture the uncertainties associated with our network predictions by constructing the associated error bars. We have summarized the steps of this new approach with respect to the nonlinear Poisson regression model in Fig. 2. For more specific details including the error-back-propagation technique associated with ANN, readers can refer to [15, 26].

# 3 Analysis

## 3.1 Simulation and Real Data Studies

As described in the previous section, a nonlinear Poisson regression model using Bayesian ANN was constructed with NETLAB toolbox [25] in MATLAB. Our analysis include variety of simulation studies and real world data sets where each of them were partitioned into a training set (80%) and a testing set (20%). For each data set we fit a linear Poisson regression model and nonlinear Poisson regression models with ANN using both ML and Bayesian approaches (HMC and hybrid Bayesian).

A 5-fold cross-validation technique was used with ML approach to minimize the network overfitting. We repeated the same process for different hyperparameter values $\boldsymbol{\alpha} = \{0.01, 0.025, 0.05, 0.075, 0.1\}$ and for different hidden nodes from 3 to 13. The final predictions are based on network committees created with ten different random initializations. When using the HMC method, we discarded some initial samples to avoid the susceptibility of sampling from a non-stationary distribution. An ARD prior (with zero mean Gaussian distribution) was used with the proposed hybrid Bayesian method. We also constructed the error bars within one standard deviation of our predictions.

In order to evaluate the model performances we used several error measurements criteria including: root mean square error (RMSE), mean absolute error (MAE), mean percentage error (MPE), and relative squared error (RSE). These error measurements help to provide an overall assessment of the predictions in different aspects. RMSE and MAE can be used to assess the prediction accuracies of the models whereas MPE acts as a good measure of bias in the predictions. RSE gives the relative error to what it would have been if a simple predictor (the average of the actual values) had been used.

We begin our review with six simulation studies. These simulation schemes are chosen in a way that the expected value of the Poisson regression model depends both linearly and nonlinearly on the covariates. For each of the above simulation schemes, we generated random samples with 500, 5000 and 50000. When evaluating the simulation studies, we have used the $\lambda_i$ instead of the actual response value $Y_i$.

- Simulation 1
  The response variable is generated with a single covariate $x \sim Uni(0, 1)$
  $$Y_i \sim Poi(\exp(x)). \tag{13}$$
- Simulation 2
  The response variable is generated with a single covariate $x \sim Uni(0, 1)$,
  $$Y_i \sim Poi(\exp(1 + 1.5\exp(x + 0.2))). \tag{14}$$
- Simulation 3
  The response variable is generated with two covariates, $x_1 \sim Uni(0, 1)$, and

$x_2 \sim Uni(0, 2)$,

$$Y_i \sim Poi(\exp(1 + 1.2\, x_1^{0.5} + 0.25\, x_2^{0.25}). \tag{15}$$

- Simulation 4

  The response variable is generated with two covariates, $x_1, x_2 \sim Uni(0, 1)$,

$$Y_i \sim Poi\left(exp\left(\frac{0.5\exp(1 + 2x_1)}{1 + \exp(x_2 + 1)}\right)\right). \tag{16}$$

- Simulation 5

  The response variable is generated with three covariates, $x_1 \sim Uni(0, 1)$, $x_2 \sim Uni(1, 2)$, *and* $x_3 \sim Uni(0, 1)$,

$$Y_i \sim Poi\left(exp\left(\frac{(0.5\, x_1^2 + x_2^2)}{1 + 0.2\exp(x_3 + 0.2)}\right)\right). \tag{17}$$

- Simulation 6

  The response variable is generated with three covariates, $x_1 \sim Uni(1, 4)$, $x_2 \sim Uni(0, 1)$, and $x_3 \sim Uni(0, 0.2)$,

$$Y_i \sim Poi(\exp(1 + 1.25\log(x_1) + 0.5x_2 + 0.25x_3^2)). \tag{18}$$

Out of the many ANN models which we created with different hidden nodes and their corresponding model evaluations, tables 1 and 2 summarize the model evaluations for ANN models with 5 and 10 hidden nodes. As can be seen from these tables, in simulation 1, linear Poisson regression model has outperformed ANN models with lower RMSE and RSE values, repeatedly. We observed similar results for all other ANN models with different hidden nodes and $\alpha$ values. This confirms the fact that, a linear model is superior when there exists a simple linear relationship between the response and the covariates.

In contrast to that, when there exist nonlinear dependencies on the covariates, ANN models have outperformed the linear Poisson regression model. More specifically, Bayesian ANN models have given the smallest prediction errors compared to the ANN models constructed with the ML method. Regardless of the sample size, the hybrid Bayesian method has performed well over HMC except for few cases. We observed the same pattern for other $\alpha$ values as well. Our findings are consistent with the study [14] performed on ordinary regression models.

We then present the model evaluations with five real world data sets (Table 3). These real-world data studies include either count or rate data. In their original applications, Poisson or negative binomial regression models were used while later model was incorporated to handle the overdispersion within the data. We compared ANN model predictions along with their original model predictions relative to RMSE, MAE and MPE for testing data. From the results, we can see that, the ANN model with hybrid Bayesian methods have the lowest RMSE and RSE values for all the data sets. Surprisingly, ANN model with ML training had higher error rates compared to the conventional Poisson or negative binomial models. In fact, it was performing well over their original models only in 2 out of 5 data sets. ANN models with HMC training have shown lower error rates 60% of the time (in 3 out of 5 data sets) compared to their original models. ANN models with new hybrid Bayesian learning were better 100% of the time (in all 5 data sets) with regards to RMSE and RSE rates while they were better 60% of the time with regards to the MAE rates. This further confirms that the ANN model with new hybrid model is better in making accurate prediction on count or rate data compared to other methods.

Figure 3 depicts actual prediction variations for the horse shoe crab testing data. We have compared the actual vs predicted outcomes using the negative binomial, ML, HMC and hybrid Bayesian ANN models. The error bars indicate the predictions within one standard deviation from the predicted mean. Ideally, this graph should illustrate a linearly increasing pattern if the model predictions matches with the actual outcomes (note that there are no crabs with 2,5 and 6 number of satellites in the testing data). Obviously, none of the current model predictions do not match with the ideal expectation. However, out of all the methods, ANN model predictions with hybrid Bayesian shows a slightly linear pattern. It has the narrowest error bars indicating reliable predictions. Predictions made by other two models fluctuate significantly along with wider error bars. These observations clearly support that the new hybrid Bayesian ANN model has the capacity of making accurate prediction for count or rate data.

Table 4 shows the relative importance of the three covariates for the horse shoe crab data set obtained using the $\alpha$ values with the ARD prior in hybrid Bayesian ANN model. These $\alpha$ values are obtained after optimizing them during the training phase of the ANN model. A lower $\alpha$ indicates a higher relative importance of that covariate to the model predictions. P-values are obtained using the corresponding negative binomial model, where a lower p-value indicate a higher importance to the predictions. The relative importance identification for each covariate either using ARD prior $\alpha$ values or p-values is consistent. In the presence of multicollinearity among covariates, use of ARD prior might be useful as it has the minimal effect to its $\alpha$ compared to the p-values of a Poisson or a negative binomial regression model. However, further studies are intended to be carried out to confirm this fact.

### 3.2 Convergence Diagnostic Check

When using the methods which utilized the Monte Carlo sampling, i.e., HMC and hybrid Bayesian, our goal is to generate samples out of the stationary distribution of the Markov chain. Therefore, we need to check whether the chain has converged or not. In order to check that, we first used a visualization technique where we overlaid 5 sequences of samples. Here, we assume that if a chain has converged, then it has forgotten its starting point. So, several sequences drawn from different starting points should be indistinguishable.

Figure 4, depicts the error function for the HMC and the proposed hybrid Bayesian methods after a burn-in period of 5000 samples for the simulation 6. As can be seen, the 5 different sequences drawn from different starting points of the two chains (two methods) are indistinguishable, which confirms the fact that samples are drawn from a stationary distribution of the Markov chain. Nevertheless, the hybrid Bayesian shows both a lesser variation and an error than in the HMC method.

We have calculated another convergence diagnostic test statistic called “estimated potential scale reduction” (EPSR) which was introduced by Gelman and Rubin [27] for each simulation. Conventionally, a group of sequence of samples can be accepted if their EPSR statistic falls below 1.10 for all statistic of interest including the regularized error function. Further details can be found in [25]. Table 5 summarizes the EPSR values of each weight parameter obtained using the ANN model with 5 hidden nodes for the Simulation 6. In contrast to the EPSR values associated with HMC weight parameters and the error function, most of the EPSR values associated with the hybrid Bayesian method are less than the cut-off 1.10. This indicates that 5,000 samples are not nearly enough for the chain to converge with the HMC method for this data set. However, we can see that our new proposed hybrid Bayesian method converge relatively faster than the HMC method as EPSR for that is less than 1.10. We observe similar results for other simulations.

## 4 Discussion

In this study, we present the details of a new nonlinear Poisson regression model developed using Bayesian ANN. As per our knowledge, this is the first study which has incorporated Bayesian learning in developing a nonlinear Poisson regression model using ANN. This model can also be used over negative binomial models when data exhibits an overdispersion. Our model has a significant potential to be used in interdisciplinary research, in addressing timely important problems related to count modeling such as accurate prediction of number of vehicles passing by an intersection in setting color light duration. According to our study, ANN with hybrid Bayesian model provides accurate predictions with respect to several error measurements criteria.

Usually, ANN models are good at making accurate predictions. However, they were repeatedly criticized due to its black box nature. Two of their substantial problems are; not being able to evaluate the significance of the covariates, and less reliability in the model prediction with high dependence on initial hyperparameters. By applying Bayesian ANN models to count modelling, we were able to successfully resolve the above problems as we incorporate the core properties of the evidence procedure and the hybrid Monte Carlo sampling in our new approach. In fact, we have successfully demonstrated how we have used the ARD prior in identifying the relative importance of the covariates and obtaining reliable predictions with low variances on Poisson or negative binomial regression. Moreover, the use of ARD prior helps to overcome the problems associated with p-values in conventional regression models. That is because, the optimized hyperparameter ($\alpha$) values in ANN models do not effected substantially in the presence of multicollinearity as the p-values in Poisson or negative binomial regression models. We intended to carry out further investigations in this regard.

This proposed nonlinear Poisson regression model can be very useful when handling the big data. This is because ANN is capable of implicitly detecting all the significant interactions among the predictor variables which we cannot achieve with the regular Poisson or negative binomial regression model. The largest data set we have used in this study consists of 250 data and 8 covariates. This is certainly not a large enough for to be considered as big data. In our future work, we wish to utilize our model to evaluate its performance on large data sets.

**Conflict of Interest Statement**
The authors declare that there is no conflict of interest associated with this manuscript.

## References

1. Poisson distribution, https://en.wikipedia.org/wiki/Poisson-regression
2. Chung C.F., Agterberg F.P.: Poisson Regression Analysis and its Application. In: Chung C.F., Fabbri A.G., Sinding-Larsen R. (eds) Quantitative Analysis of Mineral and Energy Resources. NATO ASI Series (Series C: Mathematical and Physical Sciences), vol. 223. Springer, Dordrecht (1988).

3. David, M., Jemna, D.: Modeling The Frequency of Auto Insurance Claims by Means of Poisson and Negative Binomial Models. Scientific Annals of the Alexandra Ioan Cazu University of Iai Economic Sciences **62**(2), 151-68 (2015). (Online), Vol. 62(2), 2015. pp. 151-167,
4. Michener, R., Tighe, C.: A Poisson Regression Model of Highway Fatalities. The American Economic Review **82**(2), 452-56 (1992).
5. Ripley, B.D.: Pattern Recognition and Neural Networks. 1st edn. Cambridge University Press, New York (1996).
6. Bishop, C.M.: Pattern Recognition and Machine Learning, Springer (2006).
7. Ayer,T., Alagoz, O., Chhatwal, J., Shavlik, J.W., Kahn, C.E., Burnside, E.S.: Breast cancer risk estimation with artificial neural networks revisited: Discrimination and calibration. Cancer **116**(4), 3310-21 (2010).
8. Kyri, M., Cokluk, O.: Using Multinomial Logistic Regression Analysis in Artificial Neural Network: An Application. Ozean Journal of Applied Sciences **3**(2), 259-68 (2010).
9. Mathieson, M.J.: Ordinal models for neural networks. Neural networks in financial engineering, (1996).
10. Fallah, N., Gu, H., Mohammad, K., Seyyedsalehi, S. A., Nourijelyani, K., Eshraghian, M. R.: Nonlinear Poisson regression using neural networks: A simulation study. Neural Comput. Appl. **18**(8), 939-43 (2009).
11. Fornili, M., Ambrogi, F., Boracchi, P., Biganzoli, E.: Piecewise Exponential Artificial Neural Networks (PEANN) for Modeling Hazard Function with Right Censored Data. In: Formenti, E., Tagliaferri, R., Wit, E. (eds.) Computational Intelligence Methods for Bioinformatics and Biostatistics. CIBB 2013. Lecture Notes in Computer Science, vol. 8452, pp. 125-36. Springer, Cham (2014).
12. MacKay, D.J.C.: The Evidence Framework Applied to Classification Networks. Neural Comput. **4**(5), 720-36 (1992).
13. Neal, R.M.: Bayesian Learning for Neural Networks. Ph.D. thesis, University of Toronto, Canada (1994).
14. Sharaf, T.: Statistical Learning with Artificial Neural Network Applied to Health and Environmental Data. Ph.D. thesis, University of South Florida, USA (2015).
15. Rodrigo, H.: Bayesian Artificial Neural Networks in Health and Cybersecurity. Ph.D. thesis, University of South Florida, USA, (2017).
16. Floyd, C. E. , Lo, J. Y., Yun, A. J., Sullivan, D. C., Kornguth, P. J. : Prediction of breast cancer malignancy using an artificial neural network. Cancer **74**(11), 2944-8 (1994).
17. Orr, R. K. : Use of an artificial neural network to quantitate risk of malignancy for abnormal mammograms. Surgery **129**(4), 459-66 (2001).
18. Wu, Y., Giger, M. L., Doi, K., Vyborny, C. J., Schmidt, R. A., Metz, C. E. : Artificial neural networks in mammography: application to decision making in the diagnosis of breast cancer. Radiology **187**(1), 81-7 (1993).
19. MacKay, D. J. C. : Information-based objective functions for active data selection. Neural Comput. **4**, 590 (1992).
20. MacKay, D. J. C.: A practical Bayesian framework for backpropagation networks. Neural Comput. **4**(3), 448-72 (1992).
21. MacKay, D. J. C.: Bayesian methods for backpropagation networks. In: Domany E., van Hemmen

J.L., Schulten K. (eds) Models of Neural Networks III. Physics of Neural Networks, pp. 211-54. Springer, New York (1996).

22. Bishop, C.M.: Neural networks for pattern recognition. Oxford University Press, New York (1996)
23. MacKay, D. J. C.: Bayesian non-linear modelling for the 1993 energy prediction competition. In: Heidbreder, G. (ed.) Maximum Entropy and Bayesian Methods. Fundamental Theories of Physics (An International Book Series on The Fundamental Theories of Physics: Their Clarification, Development and Application), vol. 62, pp. 221-34. Springer, Dordrecht (1996).
24. Thodberg H.H.: Ace of Bayes: application of neural networks with pruning, Technical Report 1132E, The Danish Meat Research Institute, Roskilde, Denmark (1993).
25. Nabney, I.T.: Netlab- Algorithms for pattern recognition, Springer (2002).
26. Pearlmutter, B. A.: Fast Exact Multiplication by the Hessian. Neural Comput. **6**(1), 147-60 (1994).
27. Gelman, A., Rubin, D. B.: Inference from iterative simulation using multiple sequences. Statistical Science **7**(4), 457-72 (1992).

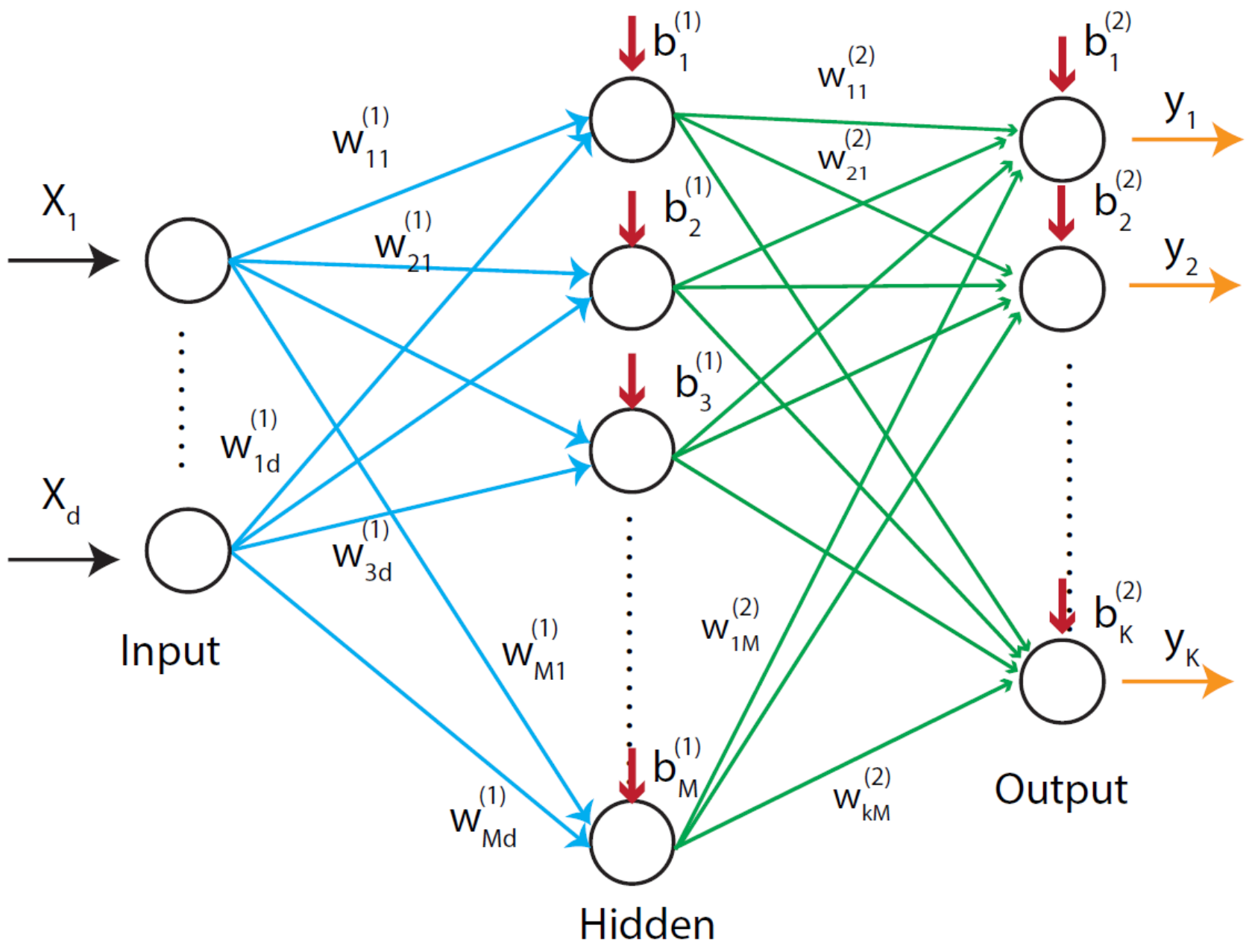

**Figure 1**. An artificial neural network with 3 layers: input, hidden and output

Step 1: Choose an initial value for the hyperparameters $\boldsymbol{\alpha}$. Initialize the weights and the bias parameters in the network.

Step 2: Train the network with a suitable optimization algorithm to minimize the regularized canonical error function *S(w)* given in equation 5.

Step 3: When the network training has achieved a local minimum, use the Gaussian approximation to compute the evidence for the hyperparameter. $\boldsymbol{\alpha}$ can be re-estimated using,

$$\boldsymbol{\alpha}^{new} = \frac{\gamma}{2E_{\boldsymbol{w}}}$$

and obtain the optimal value $\boldsymbol{\alpha}_{MAP}$

Step 4: Having found the $\boldsymbol{\alpha}_{MAP}$ and the weights and the bias parameters, use HMC to sample from the posterior distribution of the weights to approximate the predictive distribution,

$$p(t^*|\boldsymbol{x}^*, D) = \int p(t^*|\boldsymbol{x}^*, \boldsymbol{w})p(\boldsymbol{w}|D, \boldsymbol{x})d\boldsymbol{w}$$

by the finite sum,

$$p(t^*|\boldsymbol{x}^*, D) \cong \frac{1}{N}\sum_{n=1}^{N} p(t^*|\boldsymbol{x}^*, \boldsymbol{w}_n),$$

Use equation 12 to calculate the standard errors.

Step 5: Repeat the steps 1 to 4 for random initial choices for the network weights in order to generate network committees.

Step 6: Make the predictions based on the network committees.

**Figure 2**. Steps of the new hybrid Bayesian learning procedure

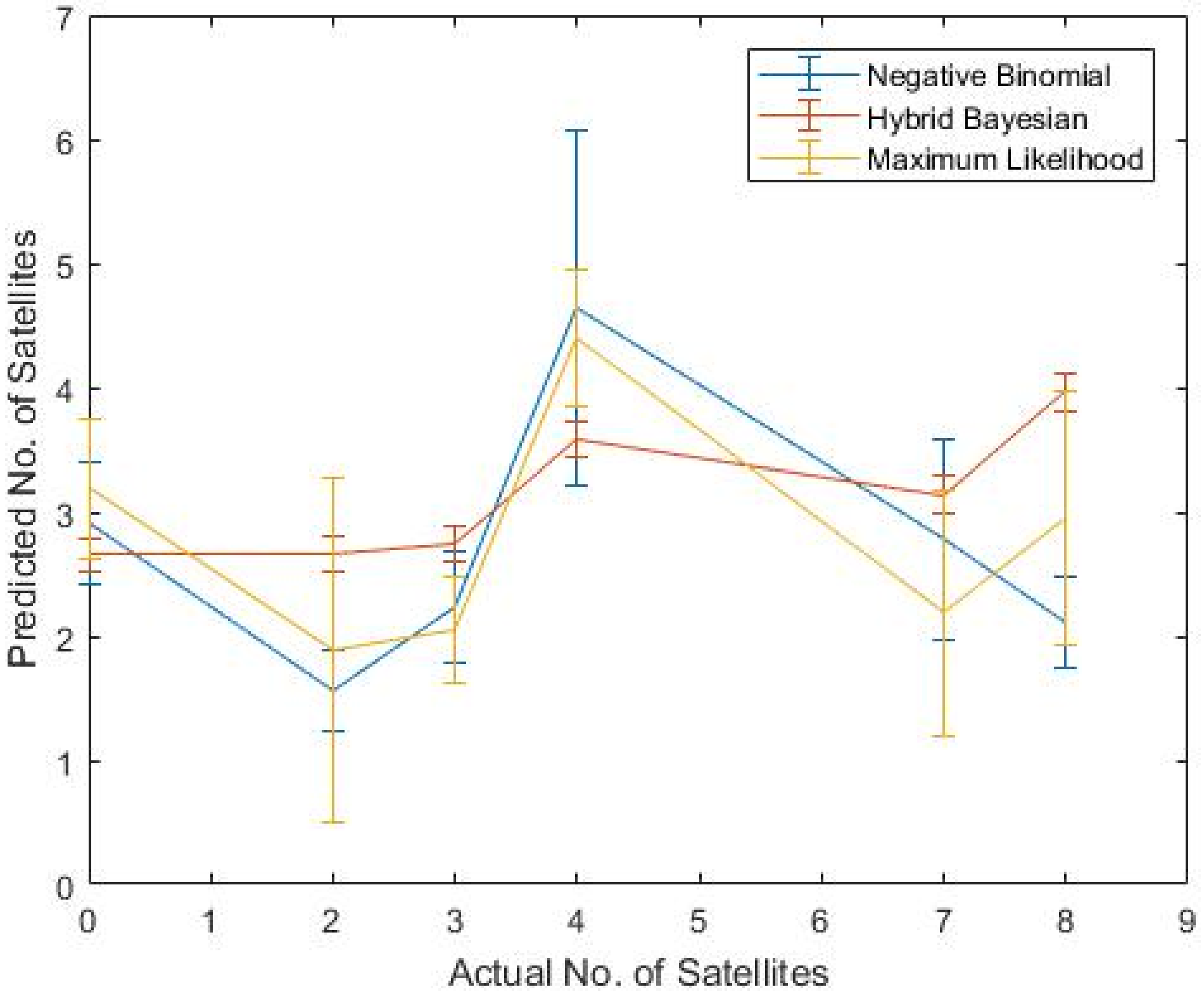

**Figure 3.** Actual vs Predicted Outcomes for Horse Shoe Crab Data

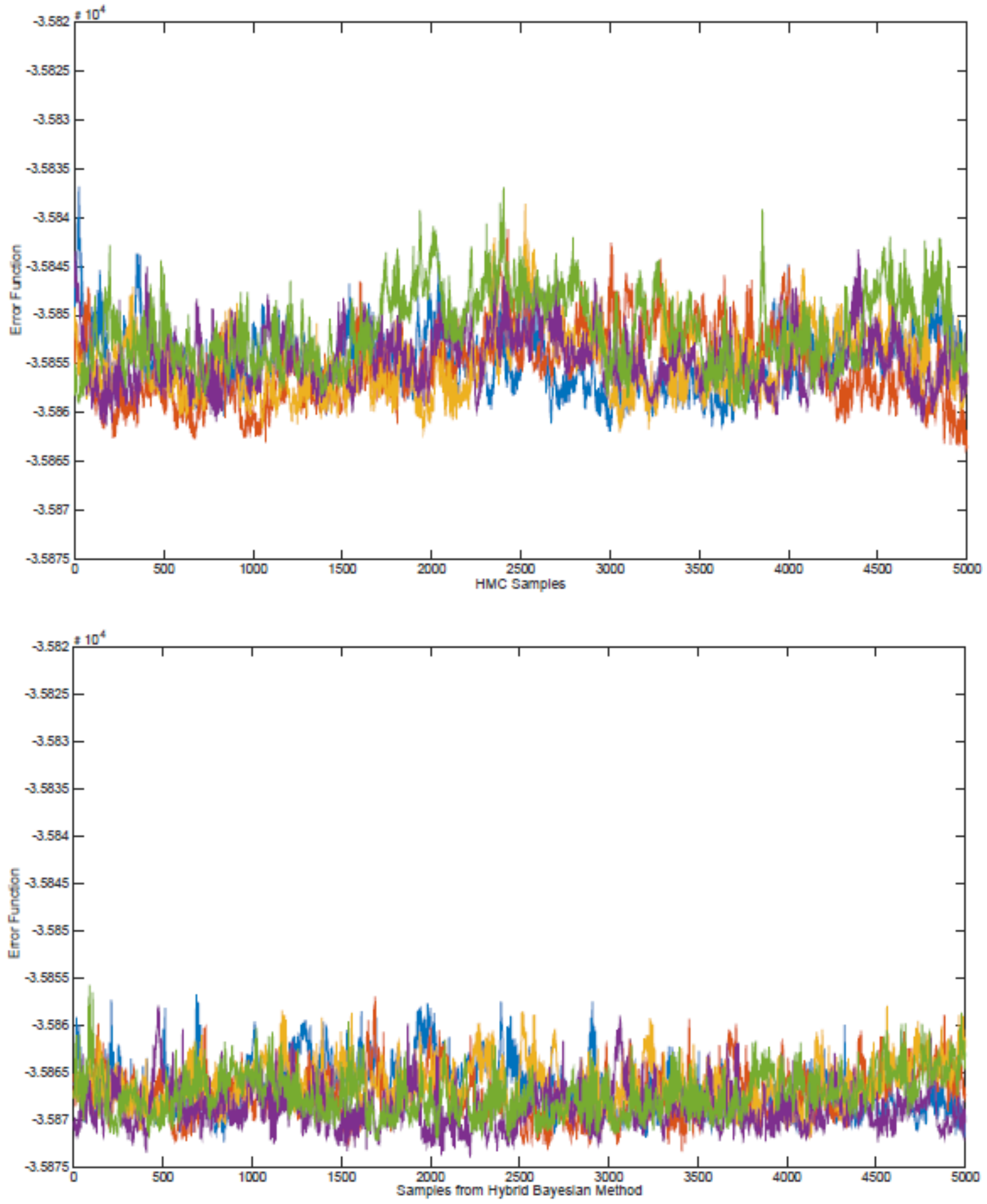

**Figure 4.** Error functions for 5 sequences drawn from the HMC and hybrid Bayesian methods after a 5000 burn-in period for simulation 6 using ANN with 5 hidden nodes

**Table 1.** Model evaluation for ANN with 5 hidden nodes with testing data. The ANN models were initiated with $\alpha = 0.075$

| | N = 500 | | | | N = 5,000 | | | | N = 50,000 | | | |
|---|---|---|---|---|---|---|---|---|---|---|---|---|
| | **RMSE** | **MAE** | **MPE** | **RSE** | **RMSE** | **MAE** | **MPE** | **RSE** | **RMSE** | **MAE** | **MPE** | **RSE** |
| ***Simulation 1*** | | | | | | | | | | | | |
| *Linear Poisson Reg* | 0.0920 | 0.0702 | 0.0367 | 0.0467 | 0.0030 | 0.0027 | 0.0019 | 0.0000 | 0.0125 | 0.0103 | 0.0061 | 0.0007 |
| *ML* | 0.2054 | 0.1541 | 0.0956 | 0.2328 | 0.1831 | 0.1604 | 0.1060 | 0.1342 | 0.1247 | 0.0656 | 0.0321 | 0.0650 |
| *HMC* | 0.0936 | 0.0688 | 0.0366 | 0.0483 | 0.0562 | 0.0462 | 0.0269 | 0.0128 | 0.0426 | 0.0351 | 0.0256 | 0.0074 |
| *Hybrid Bayesian* | 0.0924 | 0.0598 | 0.0284 | 0.0471 | 0.0328 | 0.0266 | 0.0150 | 0.0044 | 0.0169 | 0.0111 | 0.0056 | 0.0012 |
| ***Simulation 2*** | | | | | | | | | | | | |
| *Linear Poisson Reg* | 9.8679 | 7.6373 | 0.1146 | 0.0122 | 10.6772 | 7.5508 | 0.1130 | 0.1138 | 10.7062 | 7.6576 | 0.1115 | 0.0139 |
| *ML* | 2.3664 | 2.0747 | 0.0400 | 0.0006 | 1.0675 | 0.8044 | 0.0129 | 0.0001 | 0.9498 | 0.5908 | 0.0090 | 0.0001 |
| *HMC* | 0.9904 | 0.6866 | 0.0079 | 0.0001 | 0.7010 | 0.4347 | 0.0048 | 0.0001 | 0.4223 | 0.2515 | 0.0046 | 0.0000 |
| *Hybrid Bayesian* | 1.0151 | 0.7240 | 0.0095 | 0.0001 | 0.6629 | 0.4195 | 0.0048 | 0.0001 | 0.2091 | 0.1613 | 0.0031 | 0.0000 |
| ***Simulation 3*** | | | | | | | | | | | | |
| *Linear Poisson Reg* | 0.3224 | 0.2226 | 0.0437 | 0.0365 | 0.2768 | 0.2283 | 0.0389 | 0.0250 | 0.2968 | 0.2461 | 0.0428 | 0.0280 |
| *ML* | 1.0119 | 0.7661 | 0.1403 | 0.2658 | 0.2416 | 0.1957 | 0.0319 | 0.0212 | 0.0988 | 0.0823 | 0.0125 | 0.0031 |
| *HMC* | 0.2312 | 0.1868 | 0.0331 | 0.0187 | 0.1267 | 0.0971 | 0.0167 | 0.0052 | 0.0704 | 0.0466 | 0.0090 | 0.0016 |
| *Hybrid Bayesian* | 0.1898 | 0.1551 | 0.0275 | 0.0127 | 0.0997 | 0.0767 | 0.0124 | 0.0033 | 0.0388 | 0.0233 | 0.0045 | 0.0005 |
| ***Simulation 4*** | | | | | | | | | | | | |
| *Linear Poisson Reg* | 7.0242 | 5.4088 | 0.1318 | 0.0136 | 10.4293 | 5.9411 | 0.1241 | 0.0235 | 13.9159 | 7.2713 | 0.1257 | 0.0285 |
| *ML* | 5.3264 | 3.9083 | 0.0817 | 0.0079 | 1.4581 | 1.1232 | 0.0246 | 0.0005 | 1.3926 | 1.0004 | 0.0202 | 0.0003 |
| *HMC* | 1.2892 | 0.9375 | 0.0177 | 0.0005 | 1.3332 | 0.5780 | 0.0096 | 0.0004 | 0.8348 | 0.4944 | 0.0082 | 0.0001 |
| *Hybrid Bayesian* | 1.4875 | 1.0705 | 0.0203 | 0.0004 | 0.9103 | 0.4558 | 0.0096 | 0.0002 | 0.4412 | 0.3190 | 0.0063 | 0.0001 |
| ***Simulation 5*** | | | | | | | | | | | | |
| *Linear Poisson Reg* | 0.7373 | 0.4927 | 0.0876 | 0.0199 | 0.7130 | 0.4337 | 0.0646 | 0.0190 | 0.5850 | 0.4236 | 0.0720 | 0.0145 |
| *ML* | 0.6897 | 0.4105 | 0.0674 | 0.0181 | 0.7844 | 0.4881 | 0.0808 | 0.0255 | 0.3090 | 0.2053 | 0.0289 | 0.0040 |
| *HMC* | 0.5894 | 0.3540 | 0.0565 | 0.0127 | 0.4272 | 0.2337 | 0.0307 | 0.0068 | 0.2404 | 0.1605 | 0.0238 | 0.0025 |
| *Hybrid Bayesian* | 0.5873 | 0.4334 | 0.0832 | 0.0126 | 0.3264 | 0.1900 | 0.0253 | 0.0040 | 0.1965 | 0.1345 | 0.0201 | 0.0017 |
| ***Simulation 6*** | | | | | | | | | | | | |
| *Linear Poisson Reg* | 0.6915 | 0.5497 | 0.0617 | 0.0214 | 0.7422 | 0.6470 | 0.0737 | 0.0220 | 0.7415 | 0.6294 | 0.0693 | 0.0216 |
| *ML* | 1.0856 | 0.9254 | 0.0968 | 0.0448 | 0.4805 | 0.3928 | 0.0424 | 0.0091 | 0.1038 | 0.0756 | 0.0068 | 0.0004 |
| *HMC* | 0.3977 | 0.3312 | 0.0378 | 0.0071 | 0.1566 | 0.1026 | 0.0093 | 0.0010 | 0.0508 | 0.0315 | 0.0032 | 0.0001 |
| *Hybrid Bayesian* | 0.3612 | 0.2876 | 0.0348 | 0.0058 | 0.1828 | 0.1406 | 0.0144 | 0.0013 | 0.0454 | 0.0314 | 0.0032 | 0.0001 |

**Table 2.** Model evaluation for ANN with 10 hidden nodes with testing data. The ANN models were initiated with $\alpha = 0.075$

| | N = 500 | | | | N = 5,000 | | | | N = 50,000 | | | |
|---|---|---|---|---|---|---|---|---|---|---|---|---|
| | **RMSE** | **MAE** | **MPE** | **RSE** | **RMSE** | **MAE** | **MPE** | **RSE** | **RMSE** | **MAE** | **MPE** | **RSE** |
| ***Simulation 1*** | | | | | | | | | | | | |
| *Linear Poisson Reg* | 0.0896 | 0.0605 | 0.0313 | 0.0442 | 0.0030 | 0.0027 | 0.0019 | 0.0000 | 0.0125 | 0.0103 | 0.0061 | 0.0007 |
| *ML* | 0.2055 | 0.1541 | 0.0956 | 0.2328 | 0.2054 | 0.1541 | 0.0956 | 0.2327 | 0.1031 | 0.0940 | 0.0628 | 0.0432 |
| *HMC* | 0.0920 | 0.0702 | 0.0367 | 0.0467 | 0.0261 | 0.0197 | 0.0107 | 0.0028 | 0.0386 | 0.0339 | 0.0245 | 0.0216 |
| *Hybrid Bayesian* | 0.0999 | 0.0745 | 0.0375 | 0.0551 | 0.0253 | 0.0196 | 0.0108 | 0.0026 | 0.0168 | 0.0108 | 0.0053 | 0.0012 |
| ***Simulation 2*** | | | | | | | | | | | | |
| *Linear Poisson Reg* | 9.8679 | 7.6373 | 0.1146 | 0.0122 | 10.6772 | 7.5508 | 0.1130 | 0.1138 | 10.7062 | 7.6576 | 0.1115 | 0.0139 |
| *ML* | 2.2531 | 1.9506 | 0.0370 | 0.0006 | 0.6090 | 0.3415 | 0.0045 | 0.0000 | 0.4910 | 0.3991 | 0.0059 | 0.0000 |
| *HMC* | 1.0191 | 0.6916 | 0.0076 | 0.0001 | 0.8284 | 0.4389 | 0.0036 | 0.0001 | 0.2635 | 0.1888 | 0.0030 | 0.0000 |
| *Hybrid Bayesian* | 1.1441 | 0.8068 | 0.0098 | 0.0002 | 0.5702 | 0.3317 | 0.0031 | 0.0000 | 0.2402 | 0.1885 | 0.0031 | 0.0000 |
| ***Simulation 3*** | | | | | | | | | | | | |
| *Linear Poisson Reg* | 0.3224 | 0.2226 | 0.0437 | 0.0365 | 0.2768 | 0.2283 | 0.0389 | 0.0250 | 0.2968 | 0.2461 | 0.0428 | 0.0280 |
| *ML* | 0.9055 | 0.7013 | 0.1374 | 0.2135 | 0.3240 | 0.2580 | 0.0419 | 0.0378 | 0.1118 | 0.0874 | 0.0141 | 0.0040 |
| *HMC* | 0.2120 | 0.5440 | 0.0288 | 0.0158 | 0.1346 | 0.0980 | 0.0168 | 0.0059 | 0.0518 | 0.3390 | 0.0065 | 0.0009 |
| *Hybrid Bayesian* | 0.2081 | 0.1531 | 0.0300 | 0.0152 | 0.1196 | 0.0901 | 0.0149 | 0.0047 | 0.0451 | 0.0288 | 0.0055 | 0.0007 |
| ***Simulation 4*** | | | | | | | | | | | | |
| *Linear Poisson Reg* | 7.0242 | 5.4088 | 0.1318 | 0.0136 | 10.4293 | 5.9411 | 0.1241 | 0.0235 | 13.9159 | 7.2713 | 0.1257 | 0.0285 |
| *ML* | 4.9623 | 3.6819 | 0.0767 | 0.0069 | 2.0515 | 1.4670 | 0.0343 | 0.0009 | 1.7474 | 1.0824 | 0.0220 | 0.0005 |
| *HMC* | 1.6020 | 1.1538 | 0.0247 | 0.0007 | 0.9031 | 0.3898 | 0.0064 | 0.0002 | 0.6595 | 0.4266 | 0.0079 | 0.0001 |
| *Hybrid Bayesian* | 1.4149 | 0.9809 | 0.0183 | 0.0006 | 0.5540 | 0.3990 | 0.0078 | 0.0001 | 0.4669 | 0.3387 | 0.0067 | 0.0000 |
| ***Simulation 5*** | | | | | | | | | | | | |
| *Linear Poisson Reg* | 0.7373 | 0.4927 | 0.0876 | 0.0199 | 0.7130 | 0.4337 | 0.0646 | 0.0190 | 0.5850 | 0.4236 | 0.0720 | 0.0145 |
| *ML* | 1.7485 | 1.1099 | 0.2088 | 0.0854 | 1.1963 | 0.6613 | 0.1012 | 0.0590 | 0.8200 | 0.4615 | 0.0722 | 0.0292 |
| *HMC* | 0.6374 | 0.4301 | 0.0843 | 0.0149 | 0.4380 | 0.2002 | 0.0255 | 0.0072 | 0.1965 | 0.1345 | 0.0201 | 0.0017 |
| *Hybrid Bayesian* | 0.6843 | 0.5032 | 0.0970 | 0.0171 | 0.3832 | 0.1519 | 0.0183 | 0.0055 | 0.1815 | 0.1149 | 0.0178 | 0.0014 |
| ***Simulation 6*** | | | | | | | | | | | | |
| *Linear Poisson Reg* | 0.6915 | 0.5497 | 0.0617 | 0.0214 | 0.7422 | 0.6470 | 0.0737 | 0.0220 | 0.7415 | 0.6294 | 0.0693 | 0.0216 |
| *ML* | 1.2654 | 1.0494 | 0.1047 | 0.0605 | 0.4821 | 0.3915 | 0.4400 | 0.0092 | 0.1138 | 0.0851 | 0.0081 | 0.0005 |
| *HMC* | 0.4361 | 0.3725 | 0.0396 | 0.0085 | 0.1826 | 0.1147 | 0.0102 | 0.0013 | 0.0577 | 0.0363 | 0.0035 | 0.0001 |
| *Hybrid Bayesian* | 0.3658 | 0.3108 | 0.0349 | 0.0060 | 0.1366 | 0.0868 | 0.0078 | 0.0007 | 0.0467 | 0.0332 | 0.0033 | 0.0001 |

**Table 3.** Model evaluations for real world data sets. The dimension of the each data set is given by $(m \times n)$ where $m$ stands for the number of data points and $n$ stands for the number of covariates.

| | REAL WORLD DATA SETS | RMSE | MAE | RSE |
|---|---|---|---|---|
| 1 | Horse Shoe Crab Data $(173 \times 3)$ | | | |
| | Negative Binomial | 2.8968 | 2.424 | 1.1721 |
| | ML | 2.492 | 1.9945 | 0.8674 |
| | HMC | 2.4832 | 1.9915 | 0.8612 |
| | Hybrid Bayesian | 2.4686 | 1.9901 | 0.8512 |
| 2 | Student Award $(200 \times 2)$ | | | |
| | Poisson | 1.4135 | 1.0216 | 1.0018 |
| | ML | 1.8099 | 1.325 | 1.6425 |
| | HMC | 1.5863 | 1.1103 | 1.2617 |
| | Hybrid Bayesian | 1.4024 | 0.9662 | 0.9862 |
| 3 | Fish $(250 \times 3)$ | | | |
| | Negative Binomial | 2.4584 | 1.519 | 0.2076 |
| | ML | 3.3605 | 1.6876 | 0.3879 |
| | HMC | 2.355 | 1.4978 | 0.1905 |
| | Hybrid Bayesian | 2.3108 | 1.5631 | 0.1834 |
| 4 | Mussels $(42 \times 8)$ | | | |
| | Poisson | 14.7543 | 11.8949 | 2.8169 |
| | ML | 14.1962 | 11.1277 | 2.6077 |
| | HMC | 12.8026 | 9.8836 | 2.1208 |
| | Hybrid Bayesian | 10.003 | 7.5167 | 1.2947 |
| 5 | Accident $(39 \times 4)$ | | | |
| | Poisson | 1.0055 | 0.7897 | 0.8787 |
| | ML | 1.155 | 1.0461 | 1.1595 |
| | HMC | 1.1877 | 0.9921 | 1.226 |
| | Hybrid Bayesian | 0.9904 | 0.8733 | 0.8525 |

**Table 4.** Relative Covariate Importance with ARD prior with the hybrid Bayesian ANN model compared with thee the p-values of the negative binomial model

| *Covariate* | *$\alpha$ Value* | *P-Value* |
|---|---|---|
| *Width* | 83.9644 | 0.0000 |
| *Color* | 84.8513 | 0.5570 |
| *Spine* | 805.1952 | 0.9380 |

**Table 5.** Convergence diagnostics test statistic EPSR values for HMC and hybrid Bayesian methods: Simulation 6

| Weights | HMC | Hybrid Bayesian |
|---|---|---|
| w(1)_11 | 3.0594 | 1.0380 |
| w(1)_21 | 1.7617 | 1.0893 |
| w(1)_31 | 1.9997 | 1.1391 |
| w(1)_41 | 6.5150 | 1.0164 |
| w(1)_51 | 1.9991 | 1.0407 |
| w(1)_12 | 2.5444 | 1.0406 |
| w(1)_22 | 2.0649 | 1.0076 |
| w(1)_32 | 3.2607 | 1.0679 |
| w(1)_42 | 2.8220 | 1.0306 |
| w(1)_52 | 3.1978 | 1.0326 |
| w(1)_13 | 4.2560 | 1.1904 |
| w(1)_23 | 2.3057 | 1.0430 |
| w(1)_33 | 3.2699 | 1.0292 |
| w(1)_43 | 3.9502 | 1.1431 |
| w(1)_53 | 3.9652 | 1.0431 |
| b(1)_1 | 4.2990 | 2.0730 |
| b(1)_2 | 4.7093 | 1.9725 |
| b(1)_3 | 2.7830 | 1.3010 |
| b(1)_4 | 4.2292 | 1.0527 |
| b(1)_5 | 3.1815 | 1.2040 |
| w(2)_11 | 3.0967 | 2.5771 |
| w(2)_21 | 4.5723 | 1.9710 |
| w(2)_31 | 3.6035 | 1.4738 |
| w(2)_41 | 2.0472 | 1.3179 |
| w(2)_51 | 3.5783 | 2.1786 |
| b(2)_1 | 2.3792 | 1.0397 |
| Error | 1.2373 | 1.0418 |